\documentclass[times, 10pt,twocolumn]{article}
\usepackage{latex8}
\usepackage{times}

\usepackage{subfigure}
\usepackage{graphicx}
\usepackage{mathrsfs,amsmath,amsthm,color}
\usepackage{algorithm2e}
\usepackage{paralist}
\newtheorem{theorem}{{Theorem}}
\newtheorem{lemma}[theorem]{{Lemma}}

\newtheorem{definition}[theorem]{{Definition}}

\begin{document}

\title{Fountain Codes Based Distributed Storage Algorithms for  Large-scale \\Wireless Sensor Networks}

\author{Salah A. Aly\\
Dept. of Computer Science\\ Texas A\&M University\\
College Station, TX 77843\\ salah@cs.tamu.edu\\
\and
Zhenning Kong\\
Dept. of Electrical Engineering\\
Yale University\\
New Haven, CT 06520\\
zhenning.kong@yale.edu\\
\and
Emina Soljanin\\
Bell Laboratories\\
Alcatel-Lucent\\
Murray Hill, NJ 07974\\ emina@lucent.com
 }
\maketitle

\thispagestyle{empty}

\begin{abstract}
We consider large-scale networks with $n$ nodes, out of which $k$
are in possession, ({\it e.g.,} have sensed or collected in some
other way) $k$ information packets. In the scenarios in which
network nodes are vulnerable because of, for example, limited energy
or a hostile environment, it is desirable to disseminate the
acquired information throughout the network so that each of the $n$
nodes stores one (possibly coded) packet and the original $k$ source
packets can be recovered later in a computationally simple way from
any $(1 + \epsilon)k$ nodes for some small $\epsilon > 0$.

We developed two distributed algorithms for solving this problem
based on simple random walks and Fountain codes. Unlike all
previously developed schemes, our solution is truly distributed,
that is, nodes do not know $n$, $k$ or connectivity in the network,
except in their own neighborhoods, and they do not maintain any
routing tables. In the first algorithm, all the sensors have the
knowledge of $n$ and $k$. In the second algorithm, each sensor
estimates these parameters through the random walk dissemination. We
present analysis of the communication/transmission and
encoding/decoding complexity of these two algorithms, and provide
extensive simulation results as well\footnote{This work was
accomplished  while S.A.A and Z.K. were spending a summer research
internship at Bell Labs \&  Alcatel-Lucent, Murray Hill, N.J., 2007,
and it was submitted as US patent in~\cite{aly08b}. They
 would like to thank Bell Labs \&  Alcatel-Lucent staff members for their
 hospitality.}.
\end{abstract}
\section{Introduction}
Wireless sensor networks consist of small devices (sensors) with
limited resources (e.g., low CPU power, small bandwidth, limited
battery and memory). They can be deployed to monitor objects,
measure temperature, detect fires,  and other disaster phenomena.
They are often used in isolated, hard to reach areas, where human
involvement is limited. Consequently, data acquired by sensors may
have short lifetime, and any processing on it within the network
should have low complexity and power
consumption~\cite{stojmenovic05}.

We consider a large-scale wireless sensor networks with
$n$ sensors. Among them, $k\ll n$ sensors have
collected (sensed) some information. Since sensors are often short-lived
because of limited energy or hostile environment, it is
desirable to disseminate the acquired information throughout the
network so that each of the $n$ nodes stores one (possibly coded)
packet and the original $k$ source packets can be recovered in
a computationally simple way from any $(1+\epsilon)k$ of nodes for
some small $\epsilon>0$. Here, the sensors do not know locations of
each other, and they do not maintain any routing tables.

Various solutions to the centralized version of this problem have
been proposed, and are based on well known coding schemes such as
Fountain codes~\cite{dimakis06b} or MDS codes~\cite{pitkanen06}. To
distribute the information from multiple sources throughout the
network so that each node stores a coded packet as if obtained by
centralized LT (Luby Transform) coding~\cite{luby02}, Lin~\emph{et
al.}~\cite{lin07a} proposed a solution that uses random walks with
traps. To achieve the desired code degree distribution, they
employed the Metropolis algorithm to specify transition
probabilities of the random walks. In this way, the original $k$
source packets are encoded by LT codes and the decoding process can
be done by querying any $(1+\epsilon)k$ arbitrary sensors. Because
of properties of LT codes, the encoding and decoding complexity are
linear and therefore have low energy consumption.

In the methods of \cite{lin07a}, the
knowledge of the total number of sensors $n$ and sources $k$ is
required for calculating the number of random walks that each source
needs to initiate and for calculating the probability of trapping at each sensor.
Another type of global information, namely, the maximum node degree (i.e.,
the maximum number of neighbors) in the network, is also required to
perform the Metropolis algorithm. However, for a large-scale
sensor network, such global information may not be easy to
obtain by each individual sensor, especially when there is
possibility of change in topology. Moreover, the algorithms proposed
in~\cite{lin07a} assume that each sensor encodes only after receiving
enough source packets. This requires each sensor to maintain
a large enough temporary memory buffer, which may not be practical in
real sensor networks.

In this paper, we propose two new algorithms to solve the
distributed storage problem in large-scale sensor networks. We refer
to these algorithms as LT-Codes based Distributed Storage-I
(LTCDS-I) and  LT-Codes based Distributed Storage-II
(LTCDS-II). Both algorithms use simple
random walks without trapping to disseminate source packets.
In contrast to the methods in~\cite{lin07a},
both algorithms demand little global information and memory at
each sensor. In LTCDS-I, only the values of $n$ and $k$ are needed,
whereas the maximum node degree, which is more difficult to obtain, is
not required. In LTCDS-II, no sensor needs to know any
global information (that is, knowing $n$ and $k$ is no longer required). Instead,
sensors can obtain good estimates for those parameters by using some
properties of random walks. Moreover, in both algorithms, instead of
waiting until all the necessary source packets are collected to do
encoding, each sensor makes decisions and performs encoding online
upon each reception of resource packets. This mechanism reduces the
memory demand significantly.

The main contributions of this paper are as follows:
%\newpage
\begin{compactenum}[(i)] \item We propose two new algorithms
(LTCDS-I and LTCDS-II) for distributed storage in large-scale sensor
networks, using simple random walks and LT codes. These algorithms
are simpler, more robust, and less constrained in comparison to
previous solutions. \item We present complexity analysis of both
algorithms, including transmission, encoding, and decoding
complexity. \item We evaluate and illustrate the performance of both
algorithms by extensive simulation.
\end{compactenum}

This paper is organized as follows. We start with a short survey of the related
work in Section~\ref{sec:relatedwork}. In Section~\ref{sec:model_LTcodes}, we introduce the network
model and present Luby Transform (LT) codes. In
Section~\ref{sec:LTCDSalgs}, we propose two LT codes based
distributed storage algorithms called LTCDS-I and LTCDS-II.  We then
present simulation studies and provide performance analysis of the
proposed algorithms in  Section~\ref{sec:simulation},  and
concluded in Section~\ref{sec:conclusion}.

%%%%%%%%%%%%%%%%%%%%%%%%%%%%%%%%%%%%%%%%%%%%%%%%%%%%%%%%%%%%%%%%%%%%%%%%%%
\section{Related Work}\label{sec:relatedwork}
The most related work to one presented here is~\cite{lin07a,lin07b}.
Lin~\emph{el al.} studied the question ``how to retrieve historical
data that the sensors have gathered even if some sensors are
destroyed or disappeared from the network?'' They analyzed techniques
to increase {\it persistence} of sensed data in a random wireless sensor
network, and proposed two decentralized algorithms using Fountain
codes to guarantee the persistence and reliability of cached data on
unreliable sensors. They used random walks to disseminate data from
multiple sensors (sources) to the whole network. Based on the
knowledge of the total number of sensors $n$ and sources $k$, each
source calculates the number of random walks it needs to initiate,
and each sensor calculates the number of source packets it needs to
trap. In order to achieve some desired packet distribution, the
transition probabilities of random walks are specified by the well
known Metropolis algorithm~\cite{lin07a}.

Dimakis~\emph{el al.} in~\cite{dimakis06a,dimakis06b} proposed a
decentralized implementation of Fountain codes that uses geographic
routing, where every node has to know its location. The motivation
for using Fountain codes is their low decoding complexity. Also, one
does not know in advance the degrees of the output nodes in this
type of codes. The authors proposed a randomized algorithm that
constructs Fountain codes over a grid network using only
geographical knowledge of nodes and local randomized decisions. Fast
random walks are used to disseminate source data to the storage
nodes in the network.

Kamara \emph{el al.} in~\cite{kamra06,kamra05b} proposed a novel
technique called \emph{growth codes} to increase data {\it
persistence} in wireless sensor networks, namely, increase the
amount of information that can be recovered at the sink. Growth
coding is a linear technique in which information is encoded in an
online distributed way with increasing degree of a storage node.
Kamara~\emph{el al.} showed that \emph{growth codes} can increase
the amount of information that can be recovered at any storage node
at any time period whenever there is a failure in some other nodes.
They did not use robust or soliton distributions, but proposed a new
distribution depending on the network condition to determine degrees
of the storage nodes. The motivation for their work was that
\begin{inparaenum}[i)] \item Positions and topology of the nodes are not known.  \item They assume a round time
of node updates, meaning with increasing the time $t$, degree of a
symbol is increased. This is the idea behind growth degrees.
\item They provide practical implementations of growth codes and
compare its performance with other codes. \item The decoding part is
done by querying an arbitrary sink, if the original sensed data has
been collected correctly then finish, otherwise query another sink
node.
\end{inparaenum}

Lun~\emph{el. al.} in~\cite{lun05} proposed two decentralized
algorithms to compute the minimum-cost subgraphs for establishing
multicast connections using network coding. Also, they extended
their work to the problem of minimum-energy multicast in wireless
networks as well as they studied directed point-to-point multicast
and evaluated the case of elastic rate demand.

%%%%%%%%%%%%%%%%%%%%%%%%%%%%%%%%%%%%%%%%%%%%%%%%%%%%%%%%%%%%%%%%%%%%%%%%%%%%%%%
\section{Wireless Sensor Networks and Fountain Codes}\label{sec:model_LTcodes}

In this section, we introduce our network model and provide background of
Fountain codes and, in particular, one important class of Fountain codes---LT (Luby Transform)
codes~\cite{luby02}.

\subsection{Network Model}

Our wireless sensor network consists of $n$ nodes that are uniformly distributed at
random in a region $\mathcal{A}=[L,L]^2$ for $L>1$. The \emph{density} of the network is given by
\begin{equation}\label{eq:lambda}
\lambda = \frac{n}{|\mathcal{A}|}=\frac{n}{L^2},
\end{equation}
where $|\mathcal{A}|$ is the two-dimensional Lebesgue measure (or
area) of ${\cal A}$. Each sensor node has an identical communication
radius $1$; thus any two nodes can communicate with each other if
and only if their distance is less than or equal to 1. This model is
known as random geometric graphs~\cite{Gi61,Pe03}. Among these $n$
nodes, there are $k$ source nodes that have information to be
disseminated throughout the network for storage. These $k$ nodes are
uniformly and independently distributed at random among the $n$
nodes. Usually, the fraction of source nodes, i.e., $\frac{k}{n}$,
is not very large (e.g., $10\%$, or $20\%$).

Note that, although we assume the nodes are uniformly distributed at
random in a region, our algorithms and results do not rely on this
assumption. In fact, they can be applied for any network topology,
for example, regular grids.

We assume that no node has knowledge about the locations of
other nodes and no routing table is maintained; consequently, the
algorithm proposed in~\cite{DiPrRa05} cannot be applied. Moreover,
we assume that each node has limited or
no knowledge of global information, but know its neighbors. The limited global information
refers to the total numbers of nodes $n$ and sources $k$. Any
further global information, for example the maximal number of
neighbors in the network, is not available. Hence, the algorithms
proposed in~\cite{lin07a, lin07b} are not applicable.

\begin{definition}(Node Degree)
Consider a graph $G=(V,E)$, where $V$ and $E$ denote the set of nodes and
links, respectively. Given $u,v\in V$, we say $u$ and $v$ are \emph{adjacent}
(or $u$ is adjacent to $v$, and vice versa) if there exists a link between $u$
and $v$, i.e., $(u,v)\in E$. In this case, we also say that $u$ and $v$ are
\emph{neighbors}. Denote by $\mathcal{N}(u)$ the set of neighbors of a node
$u$. The number of neighbors of a node $u$ is called the \emph{node degree} of
$u$, and denoted by $d_n(u)$, i.e., $|\mathcal{N}(u)|=d_n(u)$. The \emph{mean
degree} of a graph $G$ is then given by
\begin{equation}\label{eq:mu}
\mu = \frac{1}{|V|}\sum_{u\in G}d_n(u),
\end{equation}
where $|V|$ is the total number of nodes in $G$.
\end{definition}

\subsection{Fountain Codes}
\begin{figure}[t!]\label{fig:fountaincodes}
\centering
\includegraphics[scale=0.45]{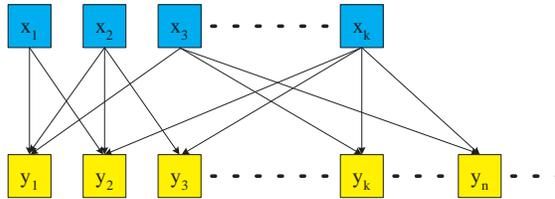}
\caption{The encoding operations of Fountain codes: each output is obtained by
XORing $d$ source blocks chosen uniformly and independently at random from $k$
source inputs, where $d$ is drawn according to a probability distribution
$\Omega(d)$.}
\end{figure}

For $k$ source blocks $\{x_1,x_2,\ldots,x_k\}$ and a probability
distribution $\Omega(d)$ with $1 \leq d \leq k$, a Fountain code
with parameters $(k,\Omega)$ is a potentially limitless stream of
output blocks $\{y_1,y_2,...\}$. Each output block is obtained by
XORing $d$ randomly and independently chosen source blocks, where
$d$ is drawn from a specially designed distribution $\Omega(d)$. This is illustrated in
Figure~\ref{fig:fountaincodes}. Fountain codes are
rateless, and one of their main advantage is that the encoding
operations can be performed online. The encoding cost is the
expected number of operation sufficient for generating an output
symbol, and the decoding cost is the expected number of operations
sufficient to recover the $k$ input blocks. Another advantage of
Fountain codes, as opposed to purely random codes is that their decoding
complexity can be made low by appropriate choice of $\Omega(d)$, with little
sacrifice in performance. The decoding of Fountain codes can be done by message passing.

\begin{definition}(Code Degree)
For Fountain codes, the number of source blocks used to generate an
encoded output $y$ is called the code degree of $y$, and denoted by
$d_c(y)$. By constraction, the code degree distribution $\Omega(d)$ is the
probability distribution of $d_c(y)$.
\end{definition}

\subsection{LT Codes}

LT (Luby Transform) codes are a special class of Fountain codes
which uses \emph{Ideal Soliton} or \emph{Robust Soliton}
distributions~\cite{luby02}. The Ideal Soliton distribution
$\Omega_{is}(d)$ for $k$ source blocks is given by
\begin{equation}\label{eq:Ideal-Soliton-distribution}
\Omega_{is}(i)=\Pr(d=i)=\left\{ \begin{array}{ll} \vspace{+.05in} \displaystyle\!\! \frac{1}{k}, &\!\! i=1\\
\displaystyle \!\!\!\frac{1}{i(i-1)}, &\!\!
i=2,3,...,k.\end{array}\right.
\end{equation}
Let $R=c_0\sqrt{k}\ln(k/\delta)$, where $c_0$ is a suitable constant
and $0<\delta<1$. The Robust Soliton distribution for $k$ source
blocks is defined as follows. Define
\begin{equation}
\tau(i)=\left\{ \begin{array}{ll} \vspace{+.05in} \displaystyle \frac{R}{ik}, & i=1,...,\displaystyle\frac{k}{R}-1\\
\vspace{+.05in} \displaystyle \frac{R\ln(R/\delta)}{k}, & i=\displaystyle\frac{k}{R}, \\
0, & i=\displaystyle\frac{k}{R}+1,...,k,\end{array}\right.
\end{equation}
and let
\begin{equation}
\beta=\sum_{i=1}^k \tau(i)+\Omega_{is}(i).
\end{equation}
The Robust Soliton distribution is given by
\begin{equation}\label{eq:Robust-Soliton-distribution}
\Omega_{rs}(i)=\frac{\tau(i)+\Omega_{is}(i)}{\beta}, \mbox{ for all }
i=1,2,...,k
\end{equation}

The following result provides the performance of the LT codes with
Robust Soliton distribution~\cite[Theorems 12 and 13]{luby02}.

\begin{lemma}[Luby~\cite{luby02}]\label{Lemma:Decoding-LT-Codes}
For LT codes with Robust Soliton distribution, $k$ original source blocks can
be recovered from any $k+O(\sqrt{k}\ln^2(k/\delta))$ encoded output blocks with
probability $1-\delta$. Both encoding and decoding complexity is
$O(k\ln(k/\delta))$.
\end{lemma}

%%%%%%%%%%%%%%%%%%%%%%%%%%%%%%%%%%%%%%%%%%%%%%%%%%%%%%%%%%%%%%%%%%%%%%%%%%%%%%
\section{LT-Codes Based Distributed Storage (LTCDS)
Algorithms}\label{sec:LTCDSalgs}

In this section, we present two LT-Codes based Distributed Storage
(LTCDS) algorithms. In both algorithms, the source packets are
disseminated throughout the network by a simple random walk. In the
first one, called LTCDS-I algorithm, we assume that each node in the
network has limited the global information, that is, knows the total number of
sources $k$ and the total number of nodes $n$. Unlike the scheme proposed in
in~\cite{lin07b}, our algorithm does not require the nodes to
know the maximum degree of the graph,
which is much harder to obtain than $k$ and $n$.
The second algorithm, called LTCDS-II, is a fully
distributed algorithm which does not require nodes to know any
global information. The price we pay for this benefit is extra
transmissions of the source packets to obtain estimates for
$n$ and $k$.

\subsection{With Limited Global Information---LTCDS-I}

In LTCDS-I, we assume that each node in the network knows the values
of $k$ and $n$. We use simple random walks~\cite{aldous02,ross95}
for each source to disseminate its information to the whole network.
At each round, each node $u$ that has packets to transmit chooses
one node $v$ among its neighbors uniformly independently at random,
and sends the packet to the node $v$. In order to avoid
local-cluster effect---each source packet is trapped most likely by
its neighbor nodes---we let each node accept a source packet
equiprobably. To achieve this, we also need each source packet to
visit each node in the network at least once.

For a random walk on a graph, the~\emph{cover time} is defined as
follows~\cite{aldous02,ross95}:
\begin{definition}(Cover Time)
Given a graph $G$, let $T_{cover}(u)$ be the expected length of a random walk
that starts at node $u$ and visits every node in $G$ at least once. The
\emph{cover time} of $G$ is defined by
\begin{equation}\label{eq:T-Cover-u}
T_{cover}(G)=\max_{u\in G}T_{cover}(u).
\end{equation}
\end{definition}

For a simple random walk on a random geometric graph, the following result
bounds the cover time~\cite{AvEr05}.

\begin{lemma}[{Avin and Ercal~\cite{AvEr05}}]\label{Lemma:Cover-Time}
If a random geometric graph with $n$ nodes is a connected graph with
high probability, then
\begin{equation}\label{eq:T-Cover-G}
T_{cover}(G)=\Theta(n\log n).
\end{equation}
\end{lemma}

As a result of Lemma \ref{Lemma:Cover-Time}, we can set a counter
for each source packet and increase the counter by one after each
forward transmission until the counter reaches some threshold
$C_1n\log n$ to guarantee that the source packet visits each node in
the network at least once. The detailed descriptions of the
initialization, encoding and storage phases (steps) of LTCDS-I
algorithm are given below:

%%%%%%%%%%%%%%%%%%%%%%%%%%%%%%%%%%%%
%\noindent \textbf{LTCDS-I Algorithm.}
\medskip

\begin{compactenum}[(i)] \item \textbf{Initialization Phase:}
\begin{compactenum}[(1)]
\item Each node $u$ in the network draws a random number $d_c(u)$ according to the distribution
$\Omega_{is}(d)$ given by~\eqref{eq:Ideal-Soliton-distribution} (or
$\Omega_{rs}(d)$ given by~\eqref{eq:Robust-Soliton-distribution}). Each source
node $s_i, i=1,\dots ,k$ generates a header for its source packet $x_{s_i}$ and
puts its ID and a counter $c(x_{s_i})$ with initial value zero into the packet
header. We set up tokens for initial and update packets. We assume that a token
is set to zero for an initial packet and $1$ for an update packet.
$$packet_{s_i}=(ID_{s_i},x_{s_i},c(x_{s_i}))$$
\item Each source node $s_i$ sends out its own source packet $x_{s_i}$ to another node $u$ which is
chosen uniformly at random among all its neighbors $\mathcal{N}(s_i)$.

\item The chosen node $u$ accepts this source $packet_{s_i}$ with probability $\frac{d_c(u)}{k}$ and updates
its storage as
    \begin{equation}
    y_u^+ = y_u^- \oplus x_{s_i},
    \end{equation}
    where $y_u^-$ and $y_u^+$ denote the packet that the node $u$ stores before and after the
    updating, respectively, and $\oplus$ represents XOR operation. No matter whether the source packet is accepted or not,
    the node $u$ puts it into its forward queue
    and set the counter of $x_{s_i}$ as
    \begin{equation}
    c(x_{s_i})=1.
    \end{equation}
\end{compactenum}
\medskip

\item  \textbf{Encoding Phase:}
\begin{compactenum}[(1)]
\item In each round, when a node $u$ receives at least one source packet before the current round,
$u$ forwards the head-of-line (HOL) packet $x$ in its forward queue to one of its neighbor $v$,
chosen uniformly at random among all its neighbors $\mathcal{N}(u)$.
\item Depending on how many times $x$ has visited $v$, the node $v$ makes its decisions:
\begin{compactenum}[\textbullet]
\item If it is the first time that $x$ visits $v$, then the node $v$ accepts this source packet
with probability $\frac{d}{k}$ and updates its storage as
    \begin{equation}\label{eq:y-u-update}
    y_v^+ = y_v^- \oplus x.
    \end{equation}
\item If $x$ has visited $v$ before and $c(x)< C_1n\log n$ where $C_1$ is a system parameter, then
the node $v$ accepts this source packet with probability 0.
\item No matter $x$ is accepted or not, the node $v$ puts it into its forward queue
    and increases the counter of $x$ by one:
    \begin{equation}
    c(x)=c(x)+1.
    \end{equation}
\item If $x$ has visited $v$ before and $c(x)\geq C_1n\log n$ then the node $v$ discards the packet
$x$ forever.
\end{compactenum}
\end{compactenum}

\medskip

\item \textbf{Storage Phase:}

When a node $u$ makes its decisions for all the source packets
$x_{s_1},x_{s_2},...,x_{s_k}$, i.e., all these packets have visited
the node $u$ at least once, the node $u$ finishes its encoding
process by declaring the current $y_u$ to be its storage packet.
\end{compactenum}
%\end{definition}
%%%%%%%%%%%%%%%%%%%%%%%%%

The pseudo-code of these steps is given in
 LTCDS-I Algorithm~\ref{alg:LTCDS-I}.

The following theorem establishes the code degree distribution of
each storage node induced by the LTCDS-I algorithm.

\begin{algorithm}[t!]
\SetLine%
\KwIn{number of nodes $n$, number of sources $k$,  source packets $x_{s_i},i=1,2,...,k$ and a positive constant $C_1$}%
\KwOut{storage packets $y_i,i=1,2,...,n$}%
\ForEach{node $u=1:n$}%
    {
    Generate $d_c(u)$ according to $\Omega_{is}(d)$ (or $\Omega_{rs}(d)$)\;%
    }%
\ForEach{source node $s_i, i=1:k$}%
    {
    Generate header of $x_{s_i}$ and $token=0$\;%
    $c(x_{s_i})=0$\;%
    Choose $u\in \mathcal{N}(s_i)$ uniformly at random, send $x_{s_i}$ to $u$\;%
    coin = rand(1)\;
    \lIf{$\mbox{coin} \leq \frac{d_c(u)}{k}$}%
        {
        $y_u$ = $y_u\oplus x_{s_i}$\;%
        }%
    Put $x_{s_i}$ into $u$'s forward queue\;%
    $c(x_{s_i})=c(x_{s_i})+1$\;%
    }%
\While{source packets remaining}%
    {
    \ForEach{node $u$ receives packets before current round}%
        {
        Choose $v\in \mathcal{N}(u)$ uniformly at random\;%
        Send HOL packet $x_{s_i}$ in $u$'s forward queue to $v$\;%
        \uIf{$v$ receives $x_{s_i}$ for the first time}%
            {
            coin = rand(1)\;
            \If{$\mbox{coin}\leq \frac{d_c(v)}{k}$}%
                {
                $y_v$ = $y_v\oplus x_{s_i}$\;%
                Put $x_{s_i}$ into $v$'s forward queue\;%
            $c(x_{s_i})=c(x_{s_i})+1$%
                }%
            }%
            \uElseIf{$c(x_{s_i})<C_1n \log n$}%
                {
                Put $x_{s_i}$ into $v$'s forward queue\;%
                $c(x_{s_i})=c(x_{s_i})+1$\;%
                }
            \Else
                {
                Discard $x_{s_i}$\;
                }%
            }%
        }%
\mbox{}\\%
\caption{LTCDS-I Algorithm: LT-Codes based Distributed Storage
Algorithm for a wireless sensor network (WSN) with limited global
information, i.e.,  values of $n$ and $k$ are known at every node.
It consists of three phases: initialization, encoding and storage
phases. The algorithm can also be deployed in a WSN after estimating
values of $n$ an $k$, as shown in LTCDS-II algorithm.}
\label{alg:LTCDS-I}
\end{algorithm}

\begin{theorem}\label{Theorem:LTCDS-I}
When a sensor network with $n$ nodes and $k$ sources finishes the storage phase
of the LTCDS-I algorithm, the code degree distribution of each storage node $u$
is given by
\begin{eqnarray}\label{eq:LTCDS-I-code-degree}
\!\!\!\!\!\!\!&&\Pr(\tilde{d}_c(u)=i) \nonumber\\
\!\!\!\!\!&=&\!\!\!\!\!\!\!\sum_{d_c(u)=1}^{k}\!\!\!
\binom{k}{i}\!\!\left(\frac{d_c(u)}{k}\right)^i\!\!\!\left(1-\frac{d_c(u)}{k}\right)^{k-i}\!\!\!\!\!\Omega'(d_c(u)),
\end{eqnarray}
where $d_c(u)$ is given in the initialization phase of the LTCDS-I algorithm
from distribution $\Omega'(d)$ (i.e., $\Omega_{is}(d)$ or $\Omega_{rs}(d)$),
and $\tilde{d}_c(u)$ is the code degree of the node $u$ resulting from the
algorithm.
\end{theorem}
\begin{proof}For each node $u$, $d_c(u)$ is drawn from a distribution $\Omega'(d)$
(i.e., $\Omega_{is}(d)$ or $\Omega_{rs}(d)$). Given $d_c(u)$, the node $u$ accepts each source
packet with probability $\frac{d_c(u)}{k}$ independently of each other and $d_c(u)$. Thus, the
number of source packets that the node $u$ accepts follows a Binomial distribution with parameter
$\frac{d_c(u)}{k}$. Hence,
\begin{eqnarray*}
\!\!\!\!\!& &\Pr(\tilde{d}_c(u)=i) \\
\!\!\!\!\!& = & \!\!\!\!\!\sum_{d_c(u)=1}^{k}\!\!\! \Pr(\tilde{d}_c(u)=i|d_c(u))\Omega'(d_c(u)\\
\!\!\!\!\!& = & \!\!\!\!\!\sum_{d_c(u)=1}^{k}\!\!\!
\binom{k}{i}\left(\frac{d_c(u)}{k}\right)^i\left(1-\frac{d_c(u)}{k}\right)^{k-i}\!\!\!\Omega'(d_c(u)),
\end{eqnarray*}
and thereafter \eqref{eq:LTCDS-I-code-degree} holds.
\end{proof}

Theorem~\ref{Theorem:LTCDS-I} indicates that the code degree
$\tilde{d}_c(u)$ is not the same as $d_c(u)$. In fact, one may
achieve the exact desired code degree distribution by letting all
the sensors hold the received source packets in their temporary
buffer until they collect all $k$ source packets. Then they can
randomly choose $d_c(u)$ packets. In this way, the resulting degree
distribution is exactly the same as $\Omega_{is}$ or $\Omega_{rs}$.
However, this requires that each sensor has enough buffer or memory,
which is usually not practical, especially when $k$ is large.
Therefore, in LTCDS-I, we assume each sensor has very limited memory
and let them make their decision upon each reception.

\begin{figure}[t!]
\centerline{ \subfigure[]{
\includegraphics[scale=0.25]{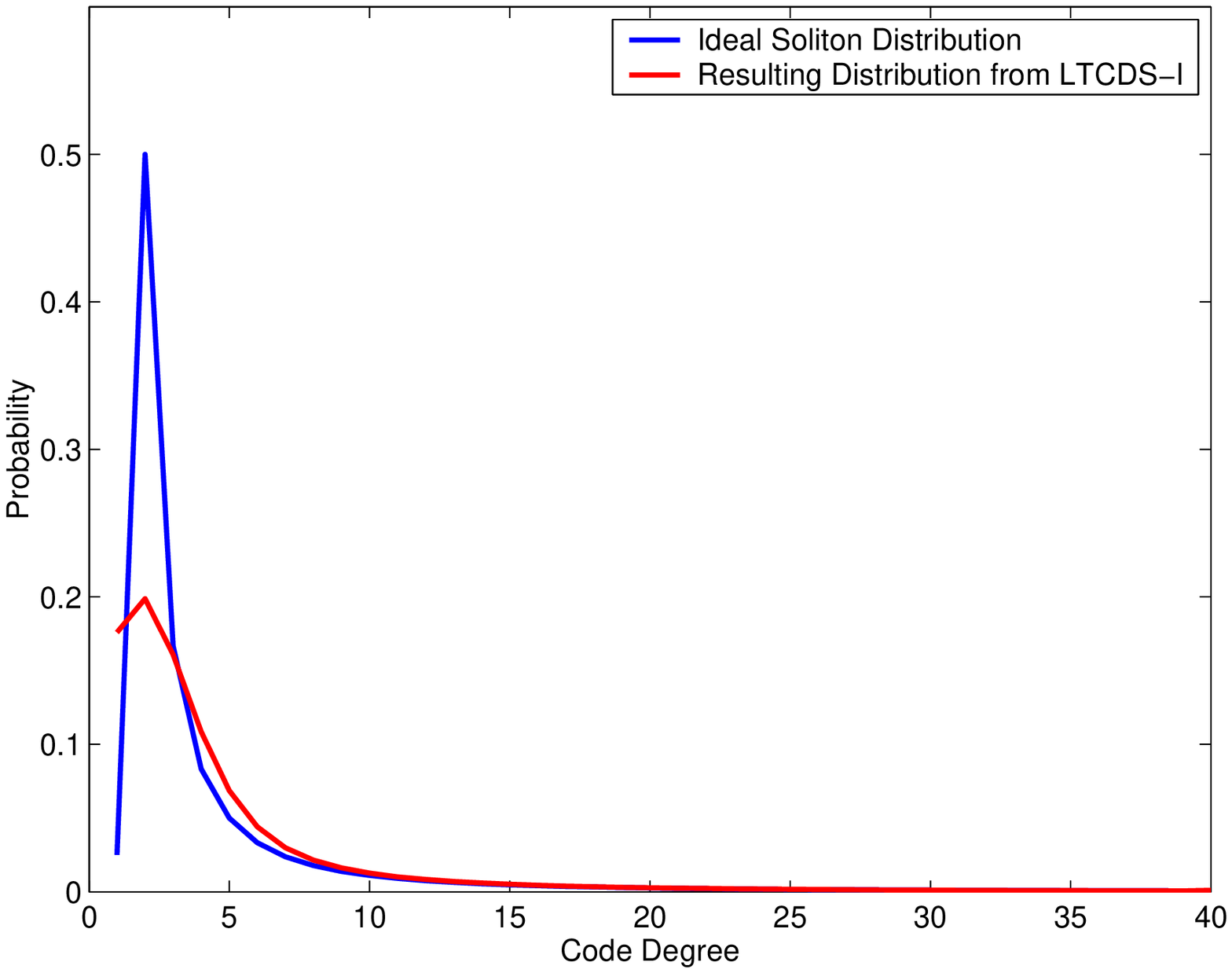}}\hfil
\subfigure[]{
\includegraphics[scale=0.25]{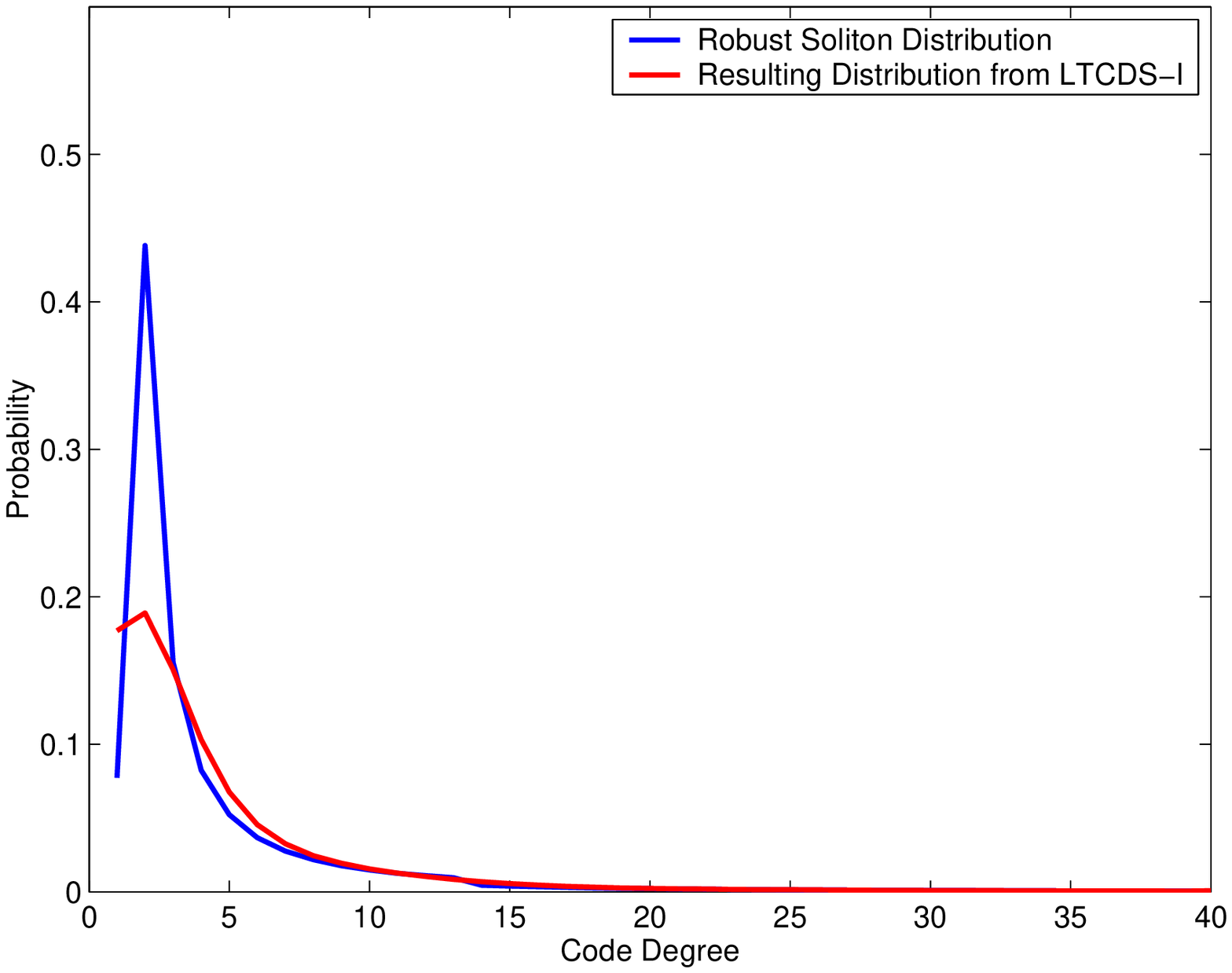}}}
\caption{Code degree distribution comparing: (a) Ideal Soliton
distribution $\Omega_{is}$ (given by
\eqref{eq:Ideal-Soliton-distribution}) and the resulting degree
distribution from LTCDS-I algorithm (given by
\eqref{eq:LTCDS-I-code-degree}). Here $k=40$; (b) Robust Soliton
distribution $\Omega_{rs}$ (given by
\eqref{eq:Robust-Soliton-distribution}) and the resulting degree
distribution from LTCDS-I algorithm (given by
\eqref{eq:LTCDS-I-code-degree}). Here $k=40$, $c_0=0.1$ and
$\delta=0.5$.}\label{fig:RSdistributino}
\end{figure}

Fortunately, from Figure~\ref{fig:RSdistributino}, we can see that
at the high degree end, the resulting code degree distribution
obtained by the LTCDS-I algorithm \eqref{eq:LTCDS-I-code-degree}
perfectly matches the desired code degree distribution, i.e., either
the Ideal Soliton distribution $\Omega_{is}$
\eqref{eq:Ideal-Soliton-distribution} or the Robust Soliton
distribution $\Omega_{rs}$ \eqref{eq:Robust-Soliton-distribution}.
For the resulting degree distribution and the desired degree
distributions, the difference only lies at the low degree end,
especially at degree 1 and degree 2. In particular, the resulting
degree distribution has higher probability at degree 1 and lower
probability at degree 2 than the desired degree distributions. The
fact that higher probability at degree 1 turns out to compensate the
lower probability at degree 2 so that the resulting degree
distribution has very similar encoding and decoding behavior as LT
codes using either the Ideal Soliton distribution or the Robust
Soliton distribution. In our future study, we will provide
theoretical analysis and prove that the degree distribution
in~\ref{eq:LTCDS-I-code-degree} is equivalent, but not the same, as
the degree distributed used in LT encoding~\cite{luby02}. Therefore,
we have the following theorem, which can be proved by the same
method for Lemma~\ref{Lemma:Decoding-LT-Codes}, see~\cite{luby02}.

\begin{theorem}\label{Theorem:Decoding-LTCDS-I}
Suppose sensor networks have $n$ nodes and $k$ sources and the LTCDS-I
algorithm uses the Robust Soliton distribution $\Omega_{rs}$. Then, when $n$
and $k$ are sufficient large, the $k$ original source packets can be recovered
from any $k+O(\sqrt{k}\ln^2(k/\delta))$ storage nodes with probability
$1-\delta$. The decoding complexity is $O(k\ln(k/\delta))$.
\end{theorem}

Theorem~\ref{Theorem:Decoding-LTCDS-I} asserts that when $n$ and $k$ are
sufficiently large, the performance of the LTCDS-I is similar to LT coding.

Another main performance metric is the transmission cost of the algorithm,
which is characterized by the total number of transmissions (the total number
of steps of $k$ random walks).

\begin{theorem}\label{Theorem:Transmission-LTCDS-I}
Denote by $T_{LTCDS}^{(I)}$ the total number of transmissions of the LTCDS-I
algorithm, then we have
\begin{equation}\label{eq:T-LTCDS-I}
T_{LTCDS}^{(I)}=\Theta(kn\log n),
\end{equation}
where $k$ is the total number of sources, and $n$ is the total number of nodes
in the network.
\end{theorem}

\begin{proof} We know that each one of $k$ source packets is stooped and discarded if and
only if it has been forwarded for $C_1n\log(n)$ times, for some
constant $C_1$. Then the total number of transmissions of the
LTCDS-I algorithm for all $k$ packets is a direct consequence and it
is given by~\eqref{eq:T-LTCDS-I}.\end{proof}

\subsection{Without any Global Information---LTCDS--II}
In many scenarios, especially when a change in network topology occurs
because of, for example, node mobility or node failures, the exact values of
$n$ and $k$ may not be available to all nodes.
Therefore, to design a fully
distributed storage algorithm which does not require any global
information is very important and useful. In this subsection, we
present such an algorithm based on LT codes, called LTCDS-II. The
idea behind this algorithm is to utilize some features of simple
random walks to do inference to obtain individual estimates of $n$
and $k$ for each node.

We  introduce of \emph{inter-visit time} and
\emph{inter-packet time}~\cite{aldous02,ross95,motwani95} as follows:
\begin{definition}(Inter-Visit Time)
For a random walk on a graph, the \emph{inter-visit time} of node $u$,
$T_{visit}(u)$, is the amount of time between any two consecutive visits of the
random walk to node $u$. This inter-visit time is also called \emph{return
time}.
\end{definition}

For a simple random walk on random geometric graphs, the following
lemma provides results on the expected inter-visit time of any node.
The proof is straightforward by following the standard result of
stationary distribution of a simple random walk on graphs and the
mean return time for a Markov
chain~\cite{aldous02,ross95,motwani95}. For completeness, we provide
the proof in Appendix~6.1.

\begin{lemma}\label{Lemma:Inter-Visit-Time}
For a node $u$ with node degree $d_n(u)$ in a random geometric graph, the mean
inter-visit time is given by
\begin{equation}\label{eq:E-T-visit-u}
E[T_{visit}(u)]=\frac{\mu n}{d_n(u)},
\end{equation}
where $\mu$ is the mean degree of the graph given by
Equation~\eqref{eq:mu}.
\end{lemma}

From Lemma~\ref{Lemma:Inter-Visit-Time}, we can see that if each node $u$ can
measure the expected inter-visit time $E[T_{visit}(u)]$, then the total number
of nodes $n$ can be estimated by
\begin{equation}
n= \frac{d_n(u)E[T_{visit}(u)]}{\mu}.
\end{equation}
However, the mean degree $\mu$ is a global information and may be hard to
obtain. Thus, we make a further approximation and let the estimate of $n$ by
the node $u$ be
\begin{equation}
\hat{n}(u) = E[T_{visit}(u)].
\end{equation}
Hence, every node $u$ computes its own estimate of $n$. In our distributed
storage algorithms, each source packet follows a simple random walk. Since
there are $k$ sources, we have $k$ individual simple random walks in the
network. For a particular random walk, the behavior of the return time is
characterized by Lemma~\ref{Lemma:Inter-Visit-Time}. On the other hand,
Lemma~\ref{Lemma:Inter-Packet-Time} below provides results on the
inter-visit time among all $k$ random walks, which is called inter-packet time
for our algorithm, defined as follows:

\begin{definition}(Inter-Packet Time)
For $k$ random walks on a graph, the \emph{inter-packet time} of node $u$,
$T_{packet}(u)$, is the amount of time between any two consecutive visits of
those $k$ random walks to node $u$.
\end{definition}

For the mean value of inter-packet time, we have the following lemma, for which
the proof is given in Appendix~6.2.

\begin{lemma}\label{Lemma:Inter-Packet-Time}
For a node $u$ with node degree $d_n(u)$ in a random geometric graph with $k$
simple random walks, the mean inter-packet time is given by
\begin{equation}\label{eq:E-T-packet-u}
E[T_{packet}(u)]=\frac{E[T_{visit}(u)]}{k}=\frac{\mu n}{kd_n(u)},
\end{equation}
where $\mu$ is the mean degree of the graph given by~\eqref{eq:mu}.
\end{lemma}

From Lemma~\ref{Lemma:Inter-Visit-Time} and
Lemma~\ref{Lemma:Inter-Packet-Time}, it is easy to see that for any node $u$,
an estimation of $k$ can be obtained by
\begin{equation}
\hat{k}(u)=\frac{E[T_{visit}(u)]}{E[T_{packet}(u)]}.
\end{equation}

After obtaining estimates for both $n$ and $k$, we can employ
similar techniques used in LTCDS-I to do LT coding and storage. The
detailed descriptions of the initialization, inference, encoding,
and storage phases of LTCDS-II algorithm are given below:

\medskip

\begin{compactenum}[(i)]
\item \textbf{Initialization Phase:}
\begin{compactenum}[(1)]
\item Each source node $s_i, i=1,\dots,k$ generates a header for its source packet $x_{s_i}$ and puts
its ID and a counter $c(x_{s_i})$ with initial value zero into the packet header.

\item Each source node $s_i$ sends out its own source packet $x_{s_i}$ to one of its neighbors $u$,
chosen uniformly at random among all its neighbors $\mathcal{N}(s_i)$.

\item The node $u$ puts $x_{s_i}$ into its forward queue and sets the counter of $x_{s_i}$ as
 \begin{eqnarray}    c(x_{s_i})=1.\end{eqnarray}
\end{compactenum}

\item \textbf{Inference Phase:}
\begin{compactenum}[(1)]
\item For each node $u$, suppose $x_{s(u)_1}$ is the first source packet that visits $u$, and
denote by $t_{s(u)_1}^{(j)}$ the time when $x_{s(u)_1}$ has its $j$-th visit to the node $u$.
Meanwhile, each node $u$ also maintains a record of visiting time for each other source packet
$x_{s(u)_i}$ that visited it. Let $t_{s(u)_i}^{(j)}$ be the time when source packet $x_{s(u)_i}$
has its $j$-th visit to the node $u$. After $x_{s(u)_1}$ visiting the node $u$ $C_2$ times, where
$C_2$ is system parameter which is a positive constant, the node $u$ stops this monitoring and
recoding procedure. Denote by $k(u)$ the number of source packets that have visited at least once
upon that time.

\item For each node $u$, let $J(s(u)_i)$ be the number of visits of source packet $x_{s(u)_i}$ to
the node $u$ and let
    \begin{eqnarray}\label{eq:T-s-u-i}
\!\!    T_{s(u)_i}\!\!&=&\!\!\!\! \frac{1}{J(s(u)_i)}\!\!\! \sum_{j=1}^{J(s(u)_i)}t_{s(u)_i}^{(j+1)}-t_{s(u)_i}^{(j)} \\
    &=&\!\!\!\! \frac{1}{J(s(u)_i)} ( t_{s(u)_i}^{(  J(s(u)_i)  )}-t_{s(u)_i}^{(1)} ).
    \end{eqnarray}
    Then, the average inter-visit time for node $u$ is given by
    \begin{equation}\label{eq:bar-T-visit-u}
    \bar{T}_{visit}(u)= \frac{1}{k(u)} \sum_{i=1}^{k(u)}T_{s(u)_i}.
    \end{equation}

    Let $J_{min}=\min_{s(u)_i}\{t_{s(u)_i}^{(1)}\}$ and $J_{max}=\max_{s(u)_i}\{t_{s(u)_i}^{(J(s(u)_i))}\}$, then
    the inter-packet time is given by
    \begin{equation}\label{eq:bar-T-packet-u}
    \bar{T}_{packet}(u)= \frac{J_{min}-J_{max}}{\sum_{s(u)_i}J(s(u)_i)}.
    \end{equation}

    Then the node $u$ can estimate the total number of nodes in the network and the total number of sources as
    \begin{equation}\label{eq:n-tilde}
    \hat{n}(u)=\bar{T}_{visit}(u),
    \end{equation}
    and
    \begin{equation}\label{eq:k-tilde}
    \hat{k}(u)=\frac{\bar{T}_{visit}(u)}{\bar{T}_{packet}(u)}.
    \end{equation}
\item In this phase, the counter $c(x_{s_i})$ of each source packet $c(x_{s_i})$ is incremented by
one after each transmission.
\end{compactenum}

\medskip

\item \textbf{Encoding Phase:}

When a node $u$ obtains estimates $\hat{n}(u)$ and $\hat{k}(u)$, it begins encoding phase which
is the same as the one in LTCDS-I Algorithm except that the code degree $d_c(u)$ is drawn from
distribution $\Omega_{is}(d)$ (or $\Omega_{rs}(d)$) with replacement of $k$ by $\hat{k}(u)$, and a
source packet $x_{s_i}$ is discarded if $c(x_{s_i})\geq C_3\hat{n}(u)\log \hat{n}(u)$, where $C_3$
is a system parameter which is a positive constant.

\medskip

\item \textbf{Storage Phase:}

When a node $u$ has made its decisions for $\hat{k}$ source packets, it finishes its
encoding process and $y_u$ becomes the storage packet of $u$.
\end{compactenum}

The total number of transmissions (the total number of steps of $k$ random
walks) in the LTCDS-II algorithm has the same order as LTCDS-I.

\begin{theorem}\label{Theorem:Transmission-LTCDS-II}
Denote by $T_{LTCDS}^{(II)}$ the total number of transmissions of the LTCDS-II
algorithm, then we have
\begin{equation}\label{eq:T-LTCDS-II}
T_{LTCDS}^{(II)}=\Theta(kn\log n),
\end{equation}
where $k$ is the total number of sources, and $n$ is the total number of nodes
in the network.
\end{theorem}

\begin{proof} In the interference phase of the LTCDS-II algorithm, the total number of
transmissions is upper bounded $C'n$ for some constants $C'>0$. That is because
each node needs to receive the first visit source packet for $C_2$ times, and
by Lemma~\ref{Lemma:Inter-Visit-Time}, the mean inter-visit time is
$\Theta(n)$.

In the decoding phase, the same as in the LTCDS-I algorithm, in
order to guarantee that each source packet visits all the nodes at
least once, the number of steps of the simple random walk is
$\Theta(n\log n)$. In other words, each source packet is stopped and
discarded if and only if the counter reaches the threshold
$C_3n\log(n)$ for some system parameter $C_3$. Therefore, we
have~\eqref{eq:T-LTCDS-II}.
\end{proof}

\medskip

\subsection{Updating Data}

Now, we turn our attention to data updating after all storage nodes
saved their values $y_1,y_2,\ldots,y_n$,  but a sensor node, say
 $s_i$, wants to update its value to the appropriate set of storage
nodes in the network. The following updating algorithm applies for
both LTCDS-I and LTCDS-II. For simplicity, we illustrate the idea
with LTCDS-I.

Assume the sensor node prepared a packet with its ID,  old data $x_{s_i}$, new data $x'_{s_i}$
along with a time-to-live parameter $c(s_i)$ initialized to zero. We will use also a simple random
walk for data update.
\begin{eqnarray}
packet_{s_i}=(ID_{s_i},x_{s_i} \oplus x'_{s_i},c(s_i)).
\end{eqnarray}

If we assume that the storage nodes keep ID's of the accepted
packets, then the problem becomes simple. We just run a random walk
and check for the coming packet's $ID$. Assume the node $u$ keeps
track of all $ID$'s of its accepted packets. Then $u$ accepts
the updated message if $ID$ of the coming packet is already included
in the $u$'s $ID$ list. Otherwise $u$ forwards the packet incrementing the
time-to-live counter. If this counter reaches the threshold value,
then the packet will be discarded.

The following steps describe the update scenario:

\medskip

\begin{compactenum}[(i)]

\item \textbf{Preparation Phase:}

The node $s_i$ prepares its new packet with the new and old data along with its ID and counter.
Also, $s_i$ add an update counter $token$ initialized at $1$ for the first updated packet. So, we
assume that the following steps happen when $token$ is set to $1$.
\begin{eqnarray}
packet_{s_i}=(ID_{s_i},x_{s_i} \oplus x'_{s_i},c(s_i)).
\end{eqnarray}
$s_i$ chooses at random a neighbor node $u$, and sends its $packet_{s_i}$.

\medskip

\item \textbf{Encoding Phase:}

The node $u$ checks if the $packet_{s_i}$ is an update or
first-time packet. If it is first-time packet it will accept,
forward, or discard it as shown in LTCDS-I
algorithm~\ref{alg:LTCDS-I}. If $packet_{s_i}$ is an updated packet,
then the node $u$ will check if $ID_{s_i}$ is already included in
its accepted list. If yes, then it will update its value $y_u$ as
follows.

\begin{eqnarray}
y_{u}^{+}=y_{u}^{-} \oplus x_{s_i}\oplus
x_{s_i}'.\end{eqnarray} If no, it will add this updated packet into
its forward queue with incrementing the counter
\begin{eqnarray}c(x_{s_i}')=c(x_{s_i}')+1.\end{eqnarray}

The $packet_{s_i}$ will be discarded if $c(x_{s_i}') \geq C_1 n \log n$ where $C_1$ is a system
parameter. In this case, we need $C_1$ to be large enough, so all old data $x_{s_i}$ will be
updated to the new data $x_{s_i}'$.

\medskip

\item \textbf{Storage Phase:}

If all nodes are done with updating their values $y_i$. One can run the decoding phase to retrieve
the original and update information.

\end{compactenum}

Now, since we run only one simple random walk for each update, if
$h$ is the number of nodes updating their values, then we have the
following result.

\begin{lemma}
The total number of transmissions needed for the update process is
bounded by $\Theta(h n \log n)$.
\end{lemma}

%%%%%%%%%%%%%%%%%%%%%%%%%%%%%%%%%%%%%%%%%%%%%%%%%%%%%%%%%%%%%%%%%%%%%%%%

\section{Performance Evaluation}\label{sec:simulation}

In this section, we study  performance of the proposed LTCDS-I and
LTCDS-II algorithms for distributed storage in wireless sensor
networks through simulation. The main performance metric we
investigate is the {\it successful decoding probability} versus the
{\it decoding ratio}.

\begin{definition}(Decoding Ratio)
\emph{Decoding ratio} $\eta$ is the ratio between the number of queried nodes
$h$ and the number of sources $k$, i.e.,
\begin{equation}\label{eq:eta}
\eta=\frac{h}{k}.
\end{equation}
\end{definition}

\begin{definition}(Successful Decoding Probability)
\emph{Successful decoding probability} $P_s$ is the probability that the $k$
source packets are all recovered from the $h$ querying nodes.
\end{definition}

In our simulation, $P_s$ is evaluated as follows. Suppose the network has $n$
nodes and $k$ sources, and we query $h$ nodes. There are $\binom{n}{h}$ ways to
choose such $h$ nodes, and we pick one tenth of these choices uniformly at random:
\begin{equation}\label{eq:M}
M=\frac{1}{10}\binom{n}{h}=\frac{n!}{10\cdot h!(n-h)!}.
\end{equation}
Let $M_s$ be the size of the subset these $M$ choices of $h$ query nodes from which the $k$ source
packets can be recovered. Then, we evaluate the successful decoding probability as
\begin{equation}\label{eq:P-s}
P_s=\frac{M_s}{M}.
\end{equation}

\begin{figure}[t!]
\centering
\includegraphics[scale=0.4]{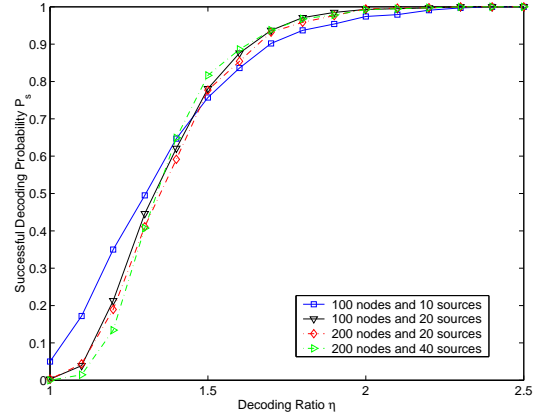}
\caption{Decoding performance of LTCDS-I algorithm with small number
of nodes and sources}\label{fig:LTCDS-I-Ps-1} % fig 3
\end{figure}

\begin{figure}[t!]
\centering
\includegraphics[scale=0.4]{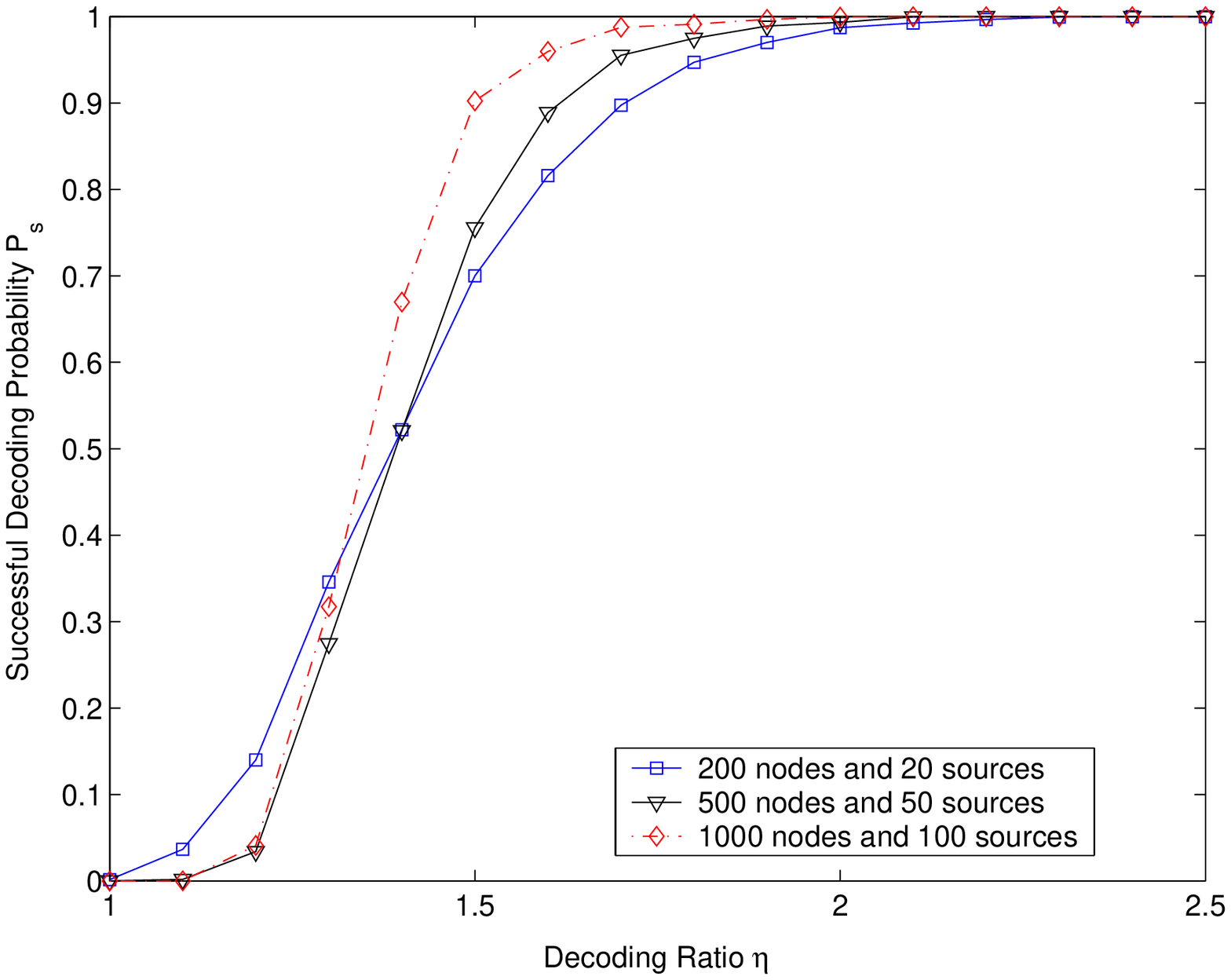}
\caption{Decoding performance of LTCDS-I algorithm with medium
number of nodes and sources}\label{fig:LTCDS-I-Ps-2} % fig 4
\end{figure}

\begin{figure}[t!]
\centering
\includegraphics[scale=0.4]{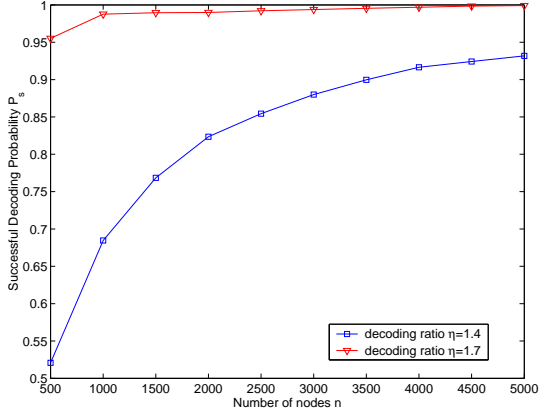}
\caption{Decoding performance of LTCDS-I algorithm with different
number of nodes}\label{fig:LTCDS-I-Ps-4} % fig 5
\end{figure}

\begin{figure}[t!]
\centering
\includegraphics[scale=0.4]{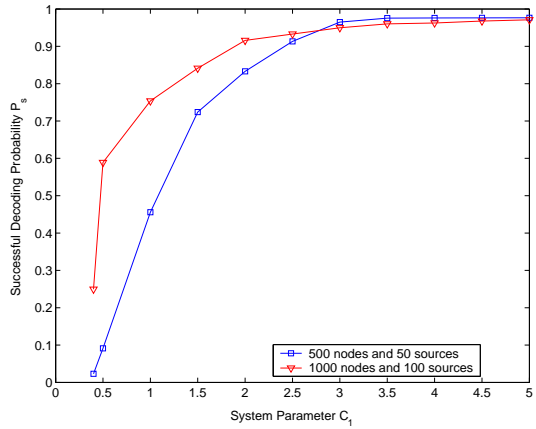}
\caption{Decoding performance of LTCDS-I algorithm with different
system parameter $C_1$}\label{fig:LTCDS-I-Ps-3} % fig 6
\end{figure}

Figure~\ref{fig:LTCDS-I-Ps-1} shows the decoding performance of
LTCDS-I algorithm with Ideal Soliton distribution with small number
of nodes and sources. The network is deployed in
$\mathcal{A}=[5,5]^2$, and the system parameter $C_1$ is set as
$C_1=5$. From the simulation results we can see that when the
decoding ratio is above 2, the successful decoding probability is
about $99\%$. Another observation is that when the total number of
nodes increases but the ratio between $k$ and $n$ and the decoding
ratio $\eta$ are kept as constants, the successful decoding
probability $P_s$ increases when $\eta\geq 1.5$ and decreases when
$\eta<1.5$. This is also confirmed by the results shown in
Figure~\ref{fig:LTCDS-I-Ps-2}. In Figure~\ref{fig:LTCDS-I-Ps-2}, The
network has constant density as $\lambda=\frac{40}{9}$ and the
system parameter $C_1=3$.

In Figure~\ref{fig:LTCDS-I-Ps-4}, we fix the decoding ratio $\eta$
as 1.4 and 1.7, respectively, and fix the ratio between the number
of sources and the number of nodes as $10\%$, i.e., $k/n=0.1$, and
change the number of nodes $n$ from 500 to 5000. From the results,
it can be seen that as $n$ grows, the successful decoding
probability increases until it reaches some platform which is the
successful decoding probability of real LT codes. This confirms that
LTCDS-I algorithm has the same asymptotical performance as LT codes.

To investigate how the system parameter $C_1$ affects the decoding
performance of the LTCDS-I algorithm, we fix the decoding ratio
$\eta$ and change $C_1$. The simulation results are shown in
Figure~\ref{fig:LTCDS-I-Ps-3}. For the scenario of 1000 nodes and
100 sources, $\eta$ is set as 1.6, and for the scenario of 500 nodes
and 50 sources, $\eta$ is set as 1.8. The code degree distribution
is also the Ideal Soliton distribution, and the network is deployed
in $\mathcal{A}=[15,15]^2$. It can be seen that when $C_1\geq 3$,
$P_s$ keeps almost like a constant, which indicates that after
$3n\log n$ steps, almost all source packets visit each node at least
once.

\begin{figure}[t!]
\centering
\includegraphics[scale=0.4]{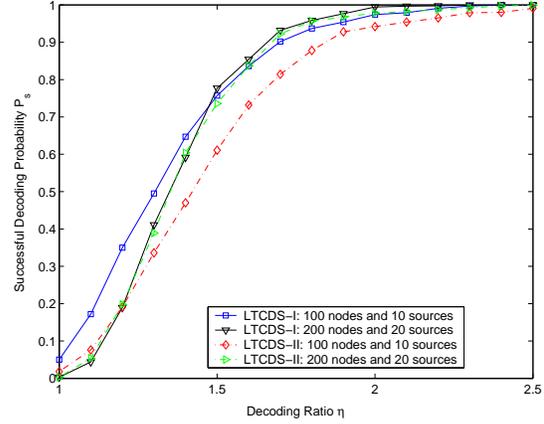}
\caption{Decoding performance of LTCDS-II algorithm with small
number of nodes and sources}\label{fig:LTCDS-II-Ps-1} %\fig 7
\end{figure}

\begin{figure}[t!]
\centering
\includegraphics[scale=0.4]{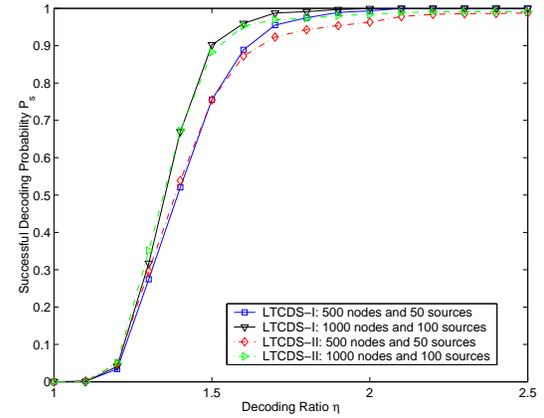}
\caption{Decoding performance of LTCDS-II algorithm with medium
number of nodes and sources}\label{fig:LTCDS-II-Ps-2} % fig 8
\end{figure}

 Figure~\ref{fig:LTCDS-II-Ps-1} compares the decoding
performance of LTCDS-II and LTCDS-I with Ideal Soliton distribution
with small number of nodes and sources. As in Figure~3, the network
is deployed in $\mathcal{A}=[5,5]^2$, and the system parameter is
set as $C_3=10$. To guarantee each node obtain accurate estimations
of $n$ and $k$, we set $C_2=50$. It can be seen that the decoding
performance of the LTCDS-II algorithm is a little bit worse than the
LTCDS-I algorithm when decoding ratio $\eta$ is small, and almost
the same when $\eta$ is large. Figure~8 compares the decoding
performance of LTCDS-II and LTCDS-I with Ideal Soliton distribution
with medium number of nodes and sources, where the network has
constant density as $\lambda=\frac{40}{9}$ and the system parameter
$C_3=20$. We observe different phenomena. The decoding performance
of the LTCDS-II algorithm is a little bit better than the LTCDS-I
algorithm when decoding ratio $\eta$ is small, and almost the same
when $\eta$ is large. That is because for the simulation in
Figure~\ref{fig:LTCDS-II-Ps-2}, we set $C_3=20$ which is larger than
$C_3=10$ set for the simulation in Figure~6. The larger value of
$C_3$ guarantees that each node has the chance to accept each source
packet, which results in a more uniformly distribution.

\begin{figure}[t!]
\centerline{ \subfigure[]{
\includegraphics[height=4cm,width=4cm]{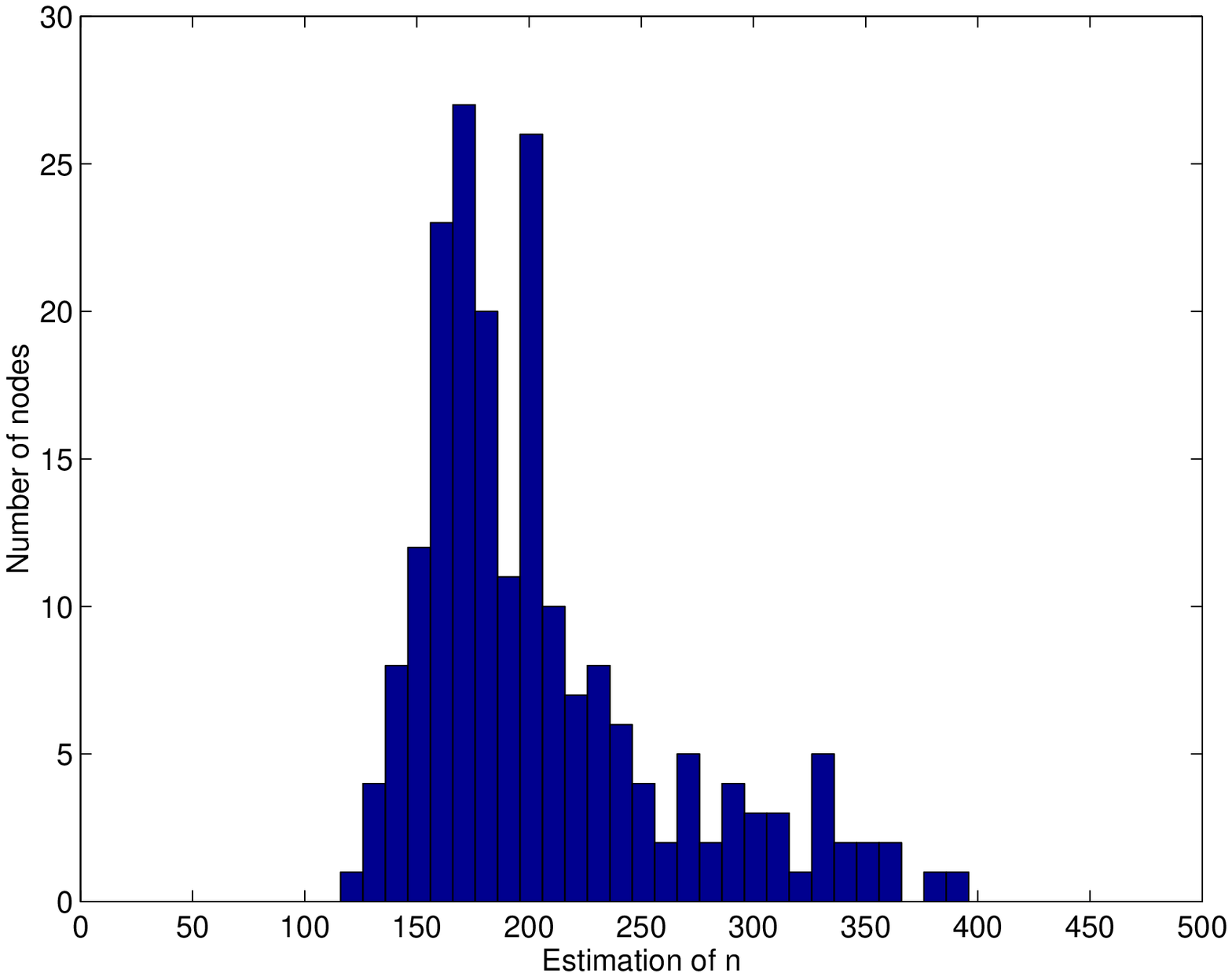}}\hfil
\subfigure[]{
\includegraphics[height=4cm,width=4cm]{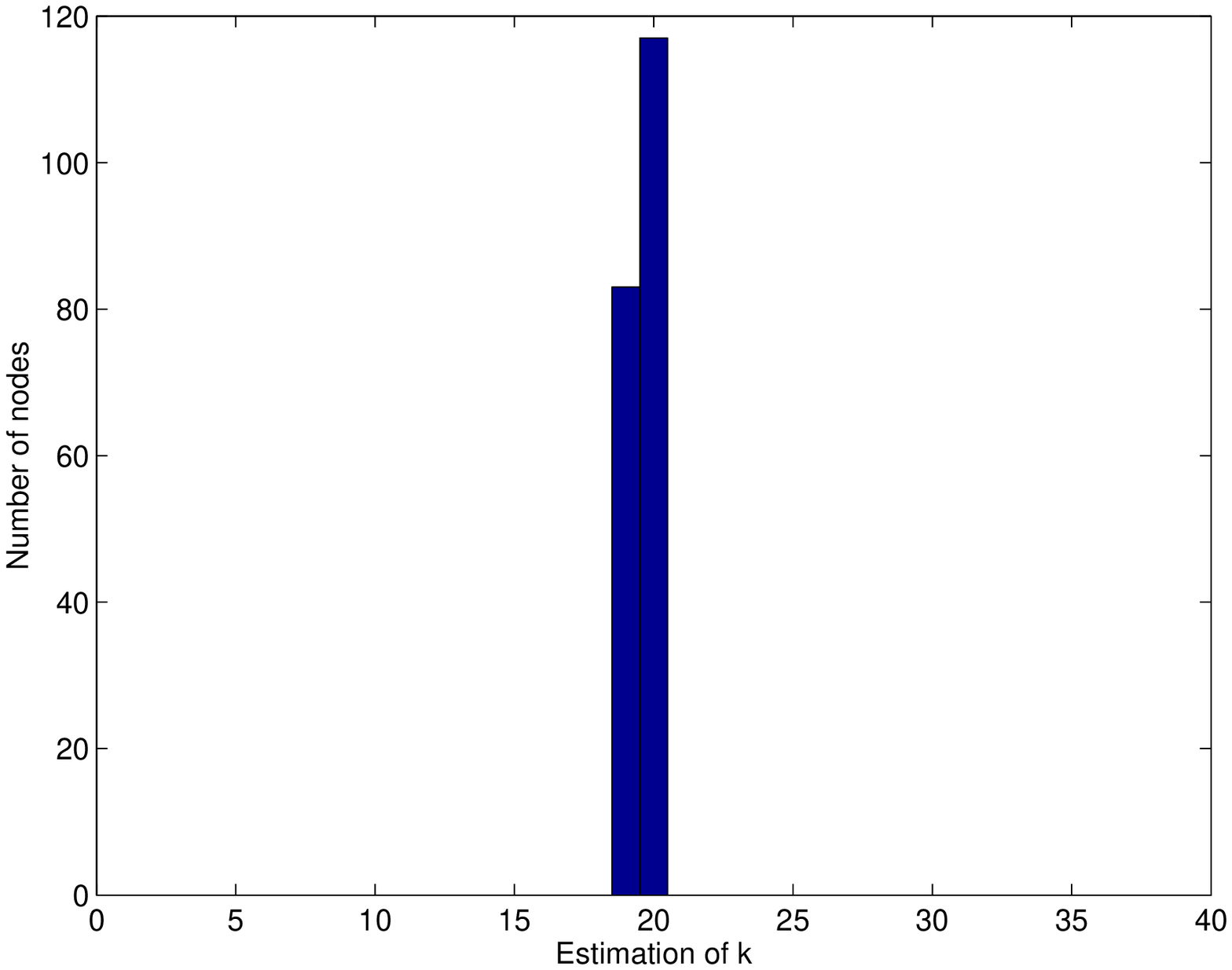}}}
\caption{Estimation results in LTCDS-II algorithm with $n=200$ nodes
and $k=20$ sources: (a) estimations of $n$; (b) estimations of $k$.
}\label{fig:LTCDS-II-Est-2} % fig 9
\end{figure}

\begin{figure}[t!]
\centerline{ \subfigure[]{
\includegraphics[height=4cm,width=4cm]{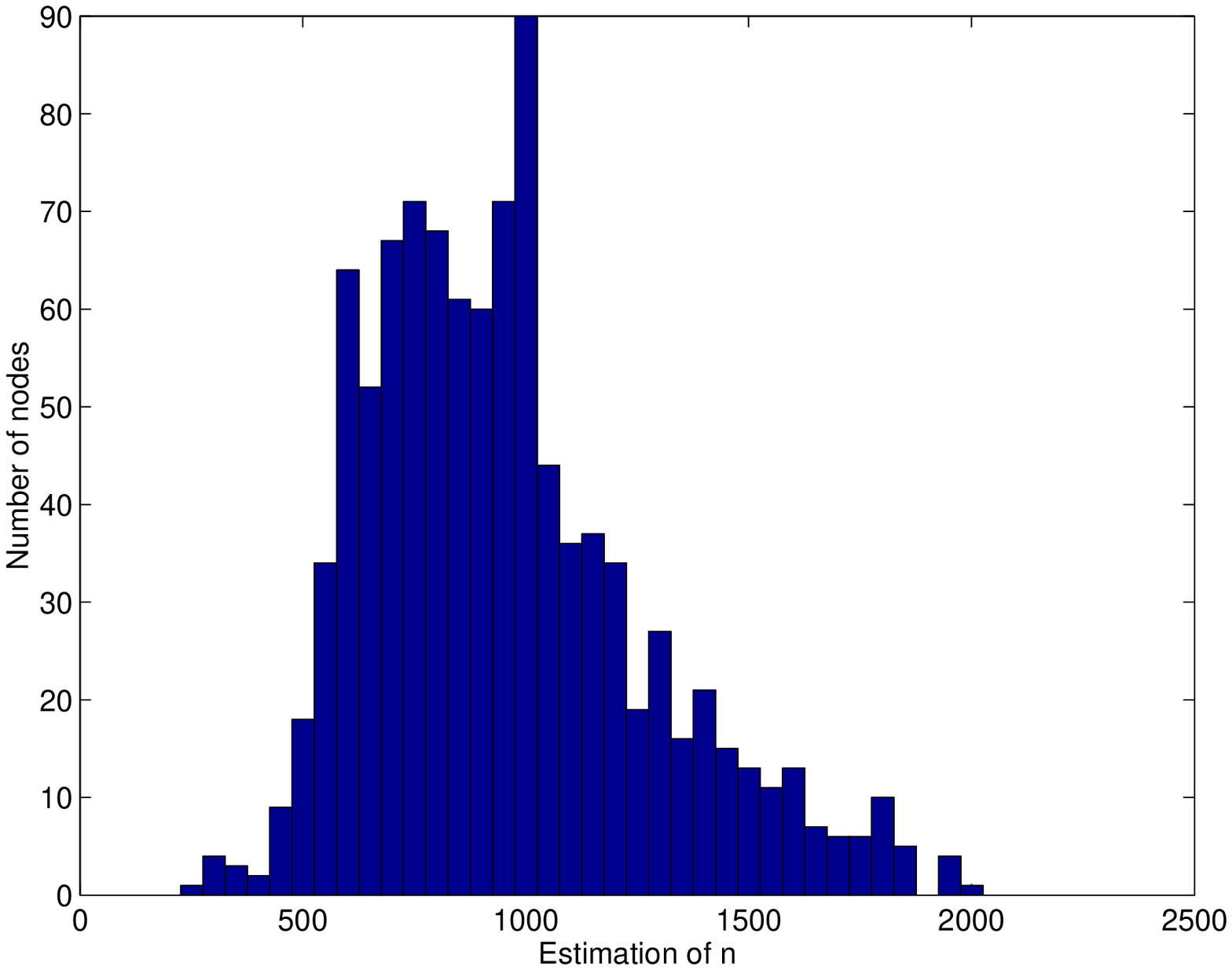}}\hfil
\subfigure[]{
\includegraphics[height=4cm,width=4cm]{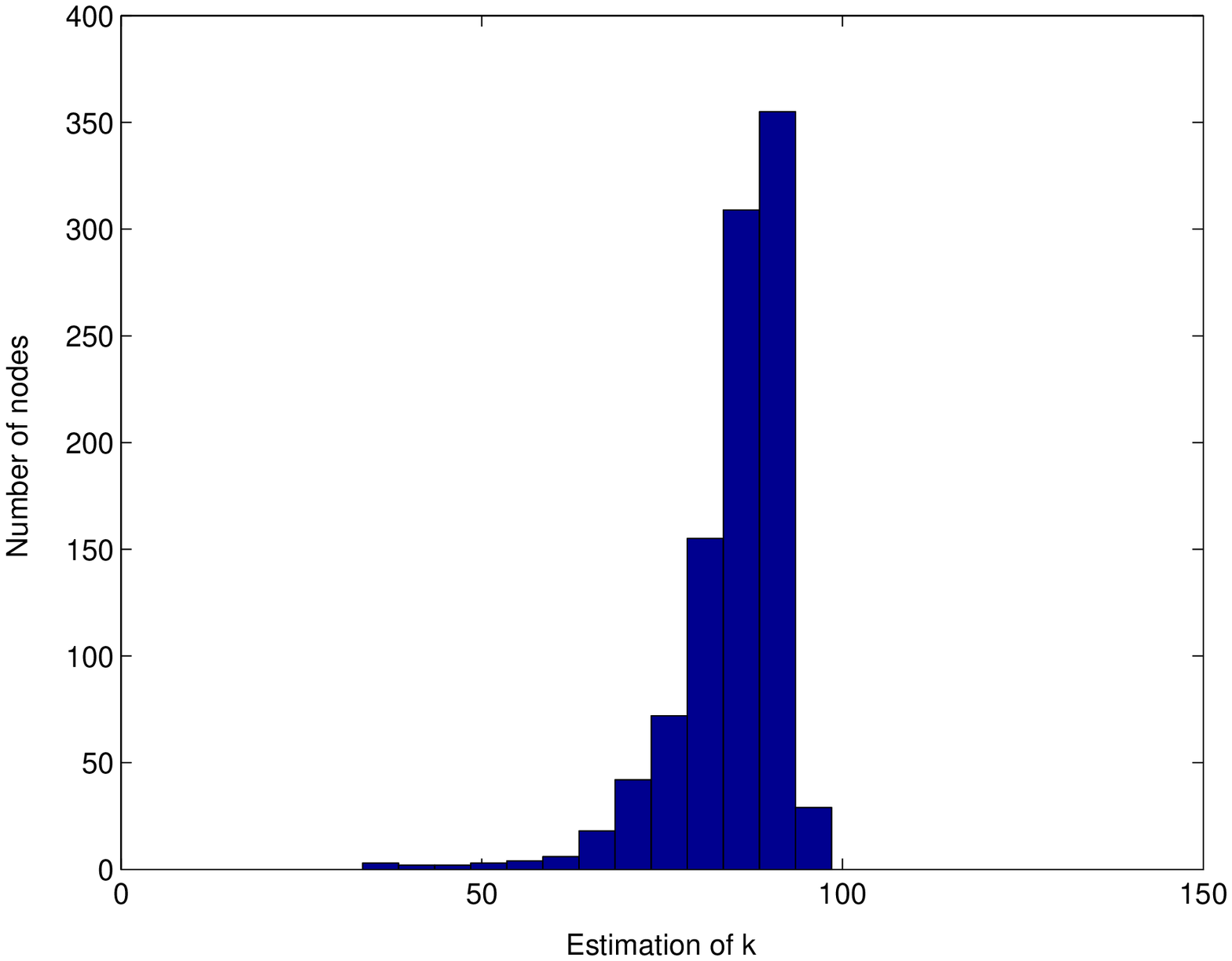}}}
\caption{Estimation results in LTCDS-II algorithm with $n=1000$
nodes and $k=100$ sources: (a) estimations of $n$; (b) estimations
of $k$.}\label{fig:LTCDS-II-Est-3} % fig 10
\end{figure}

Figure~\ref{fig:LTCDS-II-Est-2}--Figure~\ref{fig:LTCDS-II-Est-3}
shows the histogram of the estimation results of $n$ and $k$ of each
node for three scenarios: Figure~\ref{fig:LTCDS-II-Est-2} shows the
results for 200 nodes and 20 sources; and Figure~10 shows the
results for 1000 nodes and 100 sources. In the first two scenarios,
we set $C_2=50$. From the results we can see that, the estimations
of $k$ are more accurate and concentrated than the estimations of
$n$. This is because the estimation of $k$ only depends on the ratio
between the expected inter-visit time and the expected inter-packet
time, which is independent of the mean degree $\mu$ and the node
degree $d_n(u)$. On the other hand, the estimation of $n$ is
actually depends on $\mu$ and $d_n(u)$. However, in the LTCDS-II
algorithm, each node approximates $\mu$ as its own node degree
$d_n(u)$, which causes the deviation of the estimations of $n$.

\begin{figure}[t!]
\centering
\includegraphics[height=6cm]{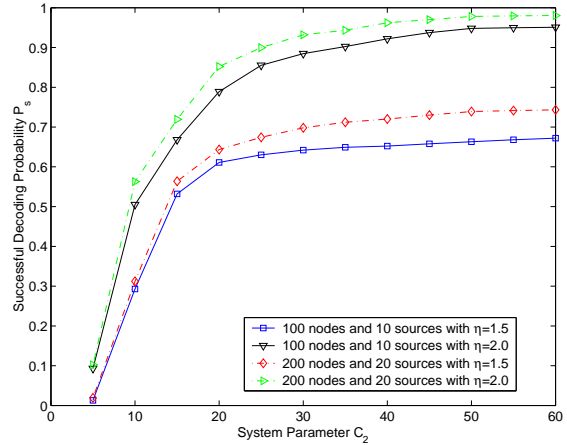}
\caption{Decoding performance of LTCDS-II algorithm with different
system parameter $C_2$}\label{fig:LTCDS-II-Ps-3} % fig 11
\end{figure}

To investigate how the system parameter $C_2$ affects the decoding
performance of the LTCDS-II algorithm, we fix the decoding ratio
$\eta$ and $C_3$, and change $C_2$. The simulation results are shown
in Figure~\ref{fig:LTCDS-II-Ps-3}. From the simulation results, we
can see that when $C_2$ is chosen to be small, the performance of
the LTCDS-II algorithm is very poor. This is due to the inaccurate
estimations of $k$ and $n$ of each node. When $C_2$ is large, for
example, when $C_2\geq 30$, the performance is almost the same.
\section{Conclusion}\label{sec:conclusion}
In this paper, we studied a model for large-scale wireless sensor
networks, where the network nodes have low CPU power and limited
storage. We proposed two new decentralized algorithms that utilize
Fountain codes and random walks to  distribute information sensed by
$k$ sensing source nodes to $n$ storage nodes. These algorithms are
simpler, more robust, and less constrained in comparison to previous
solutions that require knowledge of network topology, maximum degree
of a node, or knowing values of $n$ and
$k$~\cite{dimakis06a,dimakis06b,kamra06,lin07b,lin07a}. We computed
the computational encoding and decoding complexity of these
algorithms and simulated their performance with small and large
numbers of $k$ and $n$ nodes. We showed that a node can successfully
estimate the number of sources and total number of nodes if it can
only compute the \emph{inter-visit time} and \emph{inter-packet
time}.

Our future work will include Raptor codes based distributed
networked storage algorithms for sensor networks. We also plan to
provide theoretical results and proofs for the results shown in this
paper, where the limited space is not an issue. Our algorithm for
estimating values of $n$ and $k$ is promising, we plan to
investigate other network models where this algorithm is beneficial
and can be utilized.

\section*{Acknowledgments}
The authors would like to thank the reviewers for their comments.
 They would like to express their gratitude to all Bell Labs \&  Alcatel-Lucent staff members for their
 hospitality and kindness.
\section{Appendix}\label{sec:appendix}
\subsection{Proof of Lemma~\ref{Lemma:Inter-Visit-Time}}
\begin{proof} For a simple random walk on an undirected graph $G=(V,E)$, the stationary
distribution is given by~\cite{aldous02,ross95,motwani95}
\begin{equation}\label{eq:SRW-stationary-dsitribution}
p(u)=\frac{d_n(u)}{2|E|}.
\end{equation}

On the other hand, for a reversible Markov chain, the expected
return time for a state $i$ is given
by~\cite{aldous02,ross95,motwani95}
\begin{equation}\label{eq:MC-return-time}
E[T_{return}(i)]=\frac{1}{\pi(i)},
\end{equation}
where $\pi(i)$ is the stationary distribution of state $i$.

From~\eqref{eq:SRW-stationary-dsitribution} and~\eqref{eq:MC-return-time}, we
have for a simple random on a graph, the expected inter-visit time of node $u$
is
\begin{equation}
E[T_{visit}(u)]=\frac{2|E|}{d_n(u)}=\frac{\mu n}{d_n(u)},
\end{equation}
where $\mu$ is the mean degree of the graph. \end{proof}

\subsection{Proof of Lemma~\ref{Lemma:Inter-Packet-Time}}
\begin{proof} For a given node $u$ and $k$ simple random walks, each simple random walk
has expected inter-visit time $\frac{\mu n}{d_n(u)}$. We now view
this process from another perspective: we assume there are $k$ nodes
$\{v_1,...,v_k\}$ uniformly distributed in the network and an agent
from node $u$ follows a simple random walk. Then the expected
inter-visit time for this agent to visit any particular $v_i$ is the
same as $\frac{\mu n}{d_n(u)}$. However, the expected inter-visit
time for any two nodes $v_i$ and $v_j$ is $\frac{1}{k}\frac{\mu
n}{d_n(u)},$
which gives the expected inter-packet time.\end{proof}

\bibliographystyle{latex8}

\bibliographystyle{ieeetr}
\end{document}